\UseRawInputEncoding
\documentclass[%
 reprint,
 amsmath,amssymb,
 aps,
]{revtex4-2}

\usepackage{graphicx}
\usepackage{dcolumn}
\usepackage{bm}

\begin{document}

\preprint{APS/123-QED}

\title{Structural distortion induced Dzyaloshinskii-Moriya interaction in monolayer CrI$_{3}$ at van der Waals heterostructures}

\author{Hongxing Li}
\email{lihx@csust.edu.cn}
\author{Wei-Bing Zhang}%
\email{zhangwb@csust.edu.cn}
\affiliation{Hunan Provincial Key Laboratory of Flexible Electronic Materials Genome Engineering, School of Physics and Electronic Sciences, Changsha University of Science and Technology, Changsha 410114, People's Republic of China}%

\author{Guanghui Zhou}
\affiliation{Department of Physics, Key Laboratory for Low-Dimensional Quantum Structures and Quantum Control (Ministry of Education), and
Synergetic Innovation Center for Quantum Effects and Applications, Hunan Normal University, Changsha 410081, People's Republic of China}





\begin{abstract}
The van der Waals (vdW) magnetic heterostructures provide flexible ways to realize particular magnetic properties that possess both scientific and practical significance. Here, by first-principles calculation, we predict strong Dzyaloshinskii-Moriya interactions (DMIs) by constructing CrI$_3$/Metal vdW heterostructures. The underlaying mechanisms are ascribed the large spin-orbital coupling (SOC) of the I atom and the structural distortion in CrI$_3$ layer caused by interlayer interaction. This is different from the traditional way that deposit magnetic films on substrate to generate DMI, wherein DMI is dominated by interlayer hybridization and large SOC of substrates. In addition, both Heisenberg exchange and magnetic anisotropy are modulated dramatically, such as Heisenberg exchange is nearly doubled on Au(111), and the out-of-plane magnetism is enhanced by 88\% on Ir(111). Our work may provide a experimentally accessible strategy to induce DMI in vdW magnetic materials, which will be helpful to the design of spintronics devices.
\end{abstract}

\maketitle


\section{\label{sec:level1}Introduction}

\begin{figure*}
\begin{center}
\includegraphics[width=1\textwidth]{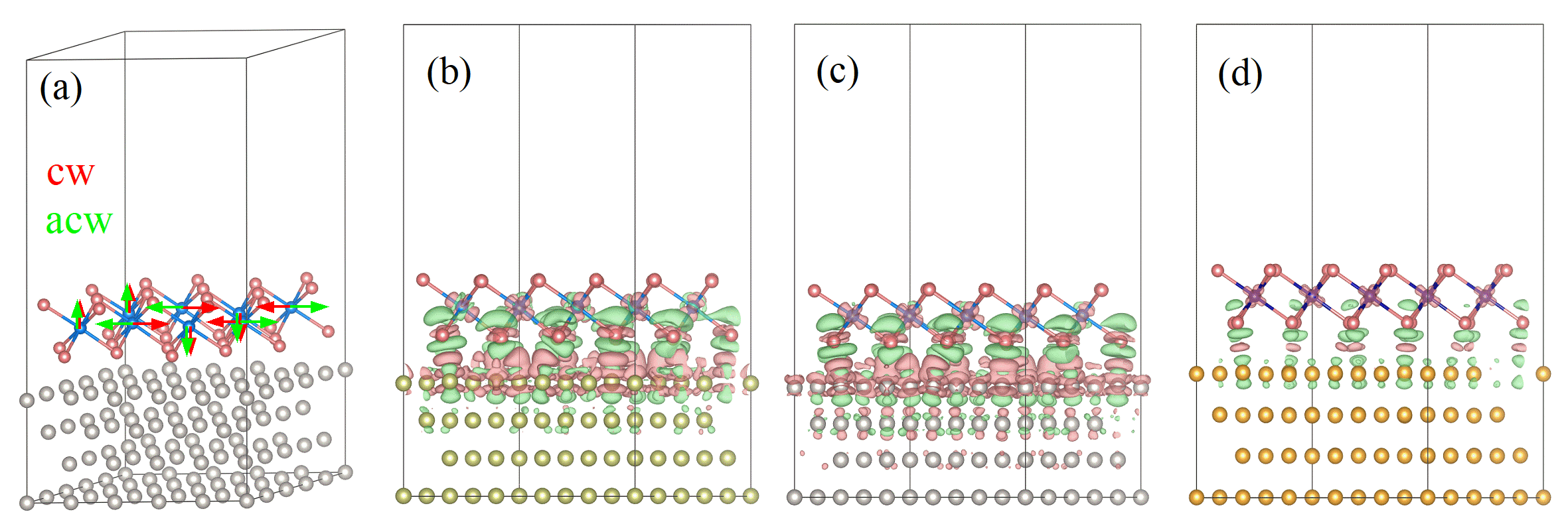}
\caption{(a) The structural schematic diagram of CrI$_3$/Metal. The red and green vectors indicate spin wave with opposite chirality. The interlayer charge density difference for (b) CrI$_3$/Ir(111), (c) CrI$_3$/Pt(111), and (d) CrI$_3$/Au(111). Pink and light green represent charge accumulation and depletion, respectively. The isosurface is set to 0.002 e/{\AA}$^3$.}
\label{Fig1}
\end{center}
\end{figure*}

Magnetism is the fundament of many modern technologies, such as magnetic storage. It is an interesting research field for a long history, and new magnetic phenomena emerge continually \cite{2022magnetism}. Due to the Pauli exclusion principle, and spin-orbital coupling (SOC) in solid, several energy terms related to magnetism are developed, including exchange interaction, magnetic anisotropy, and so on, and the magnetic properties are determined by synergetic effects of these terms \cite{TMC}. Since the magnetic energy terms are related to both atomic and electronic structures, by selecting elements and designing structures, particular magnetic properties can be achieved \cite{RMP2016}.

The magnetic interfaces are rather accessible artificial structures. In 2000, Heinze et al. \cite{heinze2000real} deposited Mn atomic monolayer on W(110) substrate, and a two-dimensional (2D) antiferromagnetic structure is revealed. In addition, because the inversion symmetry is broken in interface, Dzyaloshinskii-Moriya interaction (DMI) may arise. DMI is antisymmetric, and can result in nonlinear magnetic structure, such as Skyrmions \cite{RMP2016}. Skyrmion is a kind of spin structure that is topologically protected and behaves like particles, and can be applied in data storage and logic technologies \cite{fert2017magnetic}. The Skyrmion lattice in 2D system is firstly realized by depositing monolayer Fe film on Ir(111) surface \cite{heinze2011spontaneous}.

At present, the major strategy to generate DMI is depositing 3\emph{d} magnetic atoms on heavy 5\emph{d} nonmagnetic metal surface. Along with the experimental progress, many theoretical works have been carried out to understanding the underlying mechanisms. Belabbes et al. \cite{PRL3d_5d} systematically study a series of 3\emph{d} atoms on different 5\emph{d} substrates. They found that the strength and chirality of DMI are determined by occupations of 3\emph{d} orbital and the hybridization between 3\emph{d} and 5\emph{d} orbitals. Yang et al. \cite{PRL_Yang} elaborate that, in Co/Pt interface, despite magnetic moments are located at Cr site, the DMI is contributed by heavy Pt atoms that adjacent to Co layer. These works indicate that 3\emph{d}-5\emph{d} hybridization and the large SOC of substrates play critical roles to generate DMI.

In 2017, intrinsic 2D ferromagnetic materials CrI$_3$ and Cr$_2$Ge$_2$Te$_6$ are discovered successfully \cite{huang2017layer,gong2017discovery}, opening a new chapter in the magnetic materials. CrI$_3$ layer is ferromagnetic with out-of-plane magnetism. Both the exchange interaction and magnetic anisotropy arise from Cr-I-Cr superexchange path in which I atom has large SOC strength \cite{jmcc2016,CrI3_MAE}. However, DMI is absent in CrI$_3$ layer because the inversion symmetry. How to induce DMI in CrI$_3$ and other 2D magnetic material is an interesting topic now. Liu et al. \cite{PRB_Liu} and Behera et al. \cite{APL} theoretically induce DMI in CrI$_3$ layer by applying perpendicular electric field, but very large electric field is required. Xu et al. \cite{CrIX3} substitute one I atom layer in CrI$_3$ by Br(Cl) to fabricate Cr(I, X)$_3$ Janus monolayer, strong DMI, as well as topological spin texture are predicted.

In this study, we overlay CrI$_3$ layer on the widely-used substrates Ir(111), Pt(111) and Au(111). Note that monolayer CrI$_3$ has been prepared on Au(111) successfully \cite{li2020single}. Surprisingly, strong DMI in CrI$_3$/Metal is predicted by first-principles calculations. The primary underlaying mechanisms are ascribed to the large SOC of I atom and structural distortion in CrI$_3$ layer induced by interlayer interactions. This is different from the conventional 3\emph{d}/5\emph{d} interfaces, in which the DMI is dominated by interlayer hybridization and large SOC of substrates. Our results will provide a guideline for the design of spintronics devices based on CrI$_3$ and other 2D magnetic materials.

\section{\label{sec:level1}Calculational details}

All our calculations are performed on Vienna ab initio simulation package (VASP) based on the framework of density functional theory \cite{vasp1,vasp2}. The interaction between electron and core is calculated by projected augmented wave (PAW) method \cite{paw}. The generalized gradient approximation (GGA) with Perdew-Burke-Ernzerhof (PBE) functional \cite{gga} is adapted to evaluate the energy exchange-correlation. An effective Hubbard value U$_{eff}$=2 eV is added to Cr 3\emph{d} orbital within the scheme proposed by Dudarev et al. \cite{U}. The cutoff energy of the plane-wave basis is set to 400 eV. Brillouin zone sampling is done by a 5$\times$5$\times$1 $\Gamma$-center k-point mesh \cite{kpoint}. The convergence criteria for residual force and electronic step are 0.01 eV/{\AA} and 10$^{-5}$ eV, respectively. The DFT+D2 method is used to take the interlayer vdW force into account \cite{d2}.

The heterostructures are constructed by overlaying CrI$_3$ 2$\times$2 supercell on Ir(111), Pt(111) and Au(111) 5$\times$5 supercell with four atom layers, as shown in Fig.~\ref{Fig1}(a). During structural relaxation, the bottom atom layer of substrate is fixed to simulate the bulk condition. A vacuum slab of 15 {\AA} is added to avoid the artificial interaction between the periodic structures. The lattice constant of CrI$_3$ is optimized to be 7.01 {\AA}. To match the substrate, the CrI$_3$ monolayer should be compressed (stretched) by -3.2\%, -1.0\% and 3.0\% for Ir(111), Pt(111) and Au(111), respectively.

\section{\label{sec:level1}Results and discussions}

\begin{figure*}
\begin{center}
\includegraphics[width=0.9\textwidth]{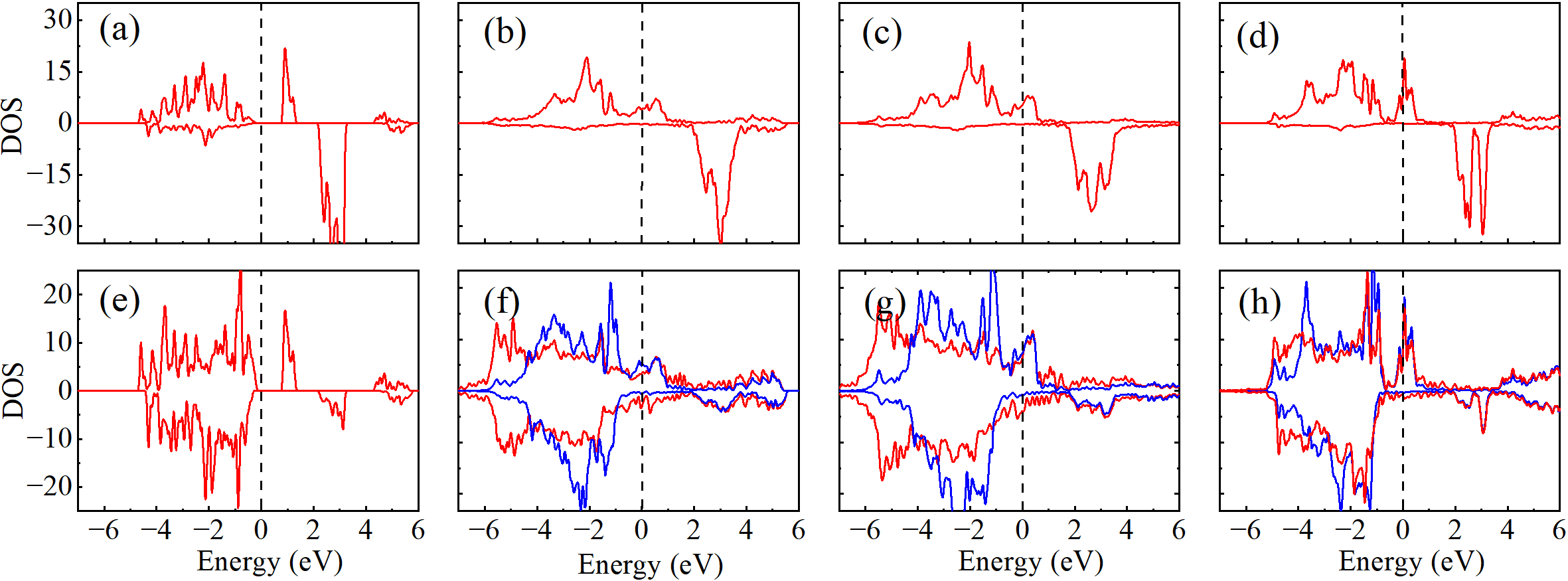}
\caption{The projected density states (PDOS) of Cr \emph{d} and I \emph{p} orbitals in CrI$_3$ layer. (a) (b) (c) and (d) The PDOS of Cr \emph{d} orbital, (e) (f) (g) and (h) the PDOS of I \emph{p} orbital for pristine CrI$_3$ monolayer, CrI$_3$/Ir(111), CrI$_3$/Pt(111) and CrI$_3$/Au(111), respectively. For PDOS of I atom, the red line and blue line represent the bottom and upper I atom layer, respectively.}
\label{Fig2}
\end{center}
\end{figure*}

\begin{table}[b]
\caption{\label{tab1}%
The calculated structural parameters for CrI$_3$/Metal heterostructures.
}
\begin{ruledtabular}
\begin{tabular}{cccccccc}
\textrm{}&
\textrm{CrI$_3$/Ir(111)}&
\textrm{CrI$_3$/Pt(111)}&
\textrm{CrI$_3$/Au(111)}\\
\colrule
\emph{d} (\AA) & 2.67 & 2.61 & 3.05 \\
Bottom Cr-I (\AA) & 2.86   & 2.81 & 2.83 \\
Upper Cr-I (\AA)  & 2.73  & 2.74 & 2.77 \\
$\Delta$\emph{l} (\AA)   & 0.13 & 0.07 & 0.06 \\
\emph{E}$_{ad}$ (meV)   & 967  & 752  & 400 \\
\end{tabular}
\end{ruledtabular}
\end{table}

After structural relaxation, the I atoms that face to substrates are pulled down, elongating the bottom Cr-I bonds. The average Cr-I bond length is summarized in Table~\ref{tab1}. We can find there is considerable difference between bottom and upper Cr-I bonds, and the largest value is 0.13 {\AA} in CrI$_3$/Ir(111), indicating remarkable structural distortion in CrI$_3$ layer. This is different from the situation that putting CrI$_3$ onto 2D substrates, which the crystal undergoes negligible deformation \cite{li2019enhanced,li2020spin,CrI3_GaN}. To evaluate the interlayer interaction, adsorption energies \emph{E}$_{ad}$ are calculated by
\begin{equation}
E_{ad}=(E_{CrI_3}+E_{Metal}-E_{CrI_3/Metal})/m
\end{equation}
where \emph{E}$_{CrI_3}$, \emph{E}$_{Metal}$ and \emph{E}$_{CrI_3/Metal}$ are the energy of CrI$_3$ layer, substrate and CrI$_3$/Metal heterostructure, and \emph{m} is number of unit cell of CrI$_3$ in CrI$_3$/Metal with a value of 4. The results are 967, 752, and 400 meV for CrI$_3$/Ir(111), CrI$_3$/Pt(111), and CrI$_3$/Au(111), respectively.

The interactions between CrI$_3$ layer and substrates will lead to charge redistributions. We calculate the charge difference density (CDD) by
\begin{equation}
\Delta\rho=\rho_{Heter}-\rho_{Metal}-\rho_{CrI_3}
\end{equation}
to further understand the interlayer interaction. The results that depicted in Fig.~\ref{Fig1} show significant charge redistribution around interface, especially CrI$_3$/Ir(111) and CrI$_3$/Pt(111), indicating strong interlayer hybridizations. On the other hand, in accordance with relative small of \emph{E}$_{ad}$, the charge redistribution in CrI$_3$/Au(111) is slight. The bottom I atoms in CrI$_3$ layer undergo great charge redistribution, while there is no obvious redistribution around the upper I atoms. Bader charge analysis \cite{bader} is used to understand the charge redistribution quantitatively. We find CrI$_3$ layer is hole-doped with 0.776 h by Pt(111), while it is electron-doped by Ir(111) and Au(111) with 0.032 e and 0.256 e, respectively. Interestingly, despite charge redistribution in CrI$_3$/Au(111) is weaker than CrI$_3$/Ir(111), there are more charge transfer in CrI$_3$/Au(111).

The average number of electrons at Cr atom, bottom and upper I atoms are summarized in Table~\ref{tab2}. In pristine CrI$_3$ layer, Cr atom is 4.687 e, and I atom is 7.438 e. Upon depositing on substrates, the amount of electrons accumulated at Cr atom range from 0.049 e to 0.138 e. On the other hand, the electrons of I atoms that face to substrates always decrease, but the upper I atom get slight electrons for all the cases. 

\begin{table}[b]
\caption{\label{tab2}%
The interlayer charge transfer $\Delta$e from substrates to CrI$_3$ layer, and number of electrons occupied at Cr and I atoms. The unit is e.}
\begin{ruledtabular}
\begin{tabular}{cccccccc}
\textrm{}&
\textrm{CrI$_3$}&
\textrm{Cr$_3$/Ir(111)}&
\textrm{Cr$_3$/Pt(111)}&
\textrm{CrI$_3$/Au(111)}\\
\colrule
 $\Delta$\emph{e} & 0 & 0.032 & -0.776 & 0.256 \\
Cr & 4.687  & 4.825 & 4.793 & 4.736 \\
Bottom I & 7.438  & 7.300 & 7.264 & 7.386 \\
Upper I   & 7.438 & 7.480 & 7.475 & 7.460 \\
\end{tabular}
\end{ruledtabular}
\end{table}

To explore the effects of interlayer interactions on the electronic structure of CrI$_3$ layer, we plot the atomic projected density of states (PDOS) in Fig.~\ref{Fig2}. For comparison, we also depict the DOS of pristine CrI$_3$ layer in Fig.~\ref{Fig2}(a) and (b). The Cr$^{3+}$ in pristine CrI$_3$ is in high spin states with 3 $\mu$$_B$ magnetic moments for three electrons occupy at spin-up t$_{2g}$ orbitals \cite{jmcc2016}. Different from the semiconductive character of pristine CrI$_3$, there are considerable states around Fermi level, thus CrI$_3$ layer undergoes metallization at interface. It should be noted that the spin splitting Cr \emph{d} bands in CrI$_3$/Au(111) is about twice as large as the pristine CrI$_3$, indicating large magnetic stability. In the case of electronic states of I atoms in CrI$_3$/Ir(111) and CrI$_3$/Pt(111), the occupied states of bottom I atom delocalizes in a wider energy range than upper one. On the other hand, due to the weak interlayer hybridization in CrI$_3$/Au(111), the PDOS of bottom and upper I atoms are still similar.

Considering the remarkable alterations in atomic and electronic structures for CrI$_3$ layer in CrI$_3$/Metal, significant modulation in magnetic properties by interfacial effects can be expected. CrI$_3$ monolayer is a well understood ferromagnet with out-of-plane magnetization. The magnetic moment (MM) of Cr atom in pristine CrI$_3$ monolayer with ferromagnetic (FM) state is 3.116 $\mu$$_B$. However, the MM increases markedly when CrI$_3$ layer is overlaid on substrates, since there are extra charges accumulate at Cr site and occupy spin-up states. On the other hand, the proximity induced MMs in substrates are several hundredth $\mu$$_B$, which are about one order smaller than the MMs polarized by 3\emph{d} magnetic atom layer \cite{PRL3d_5d}. This fact indicates the little orbitals overlap between Cr atom and substrate. This is different to 3\emph{d}/5\emph{d} interfaces, where 3\emph{d} magnetic atoms interact with substrates directly, and then polarizes the substrates to large extent.

\begin{figure}
\begin{center}
\includegraphics[width=0.5\textwidth]{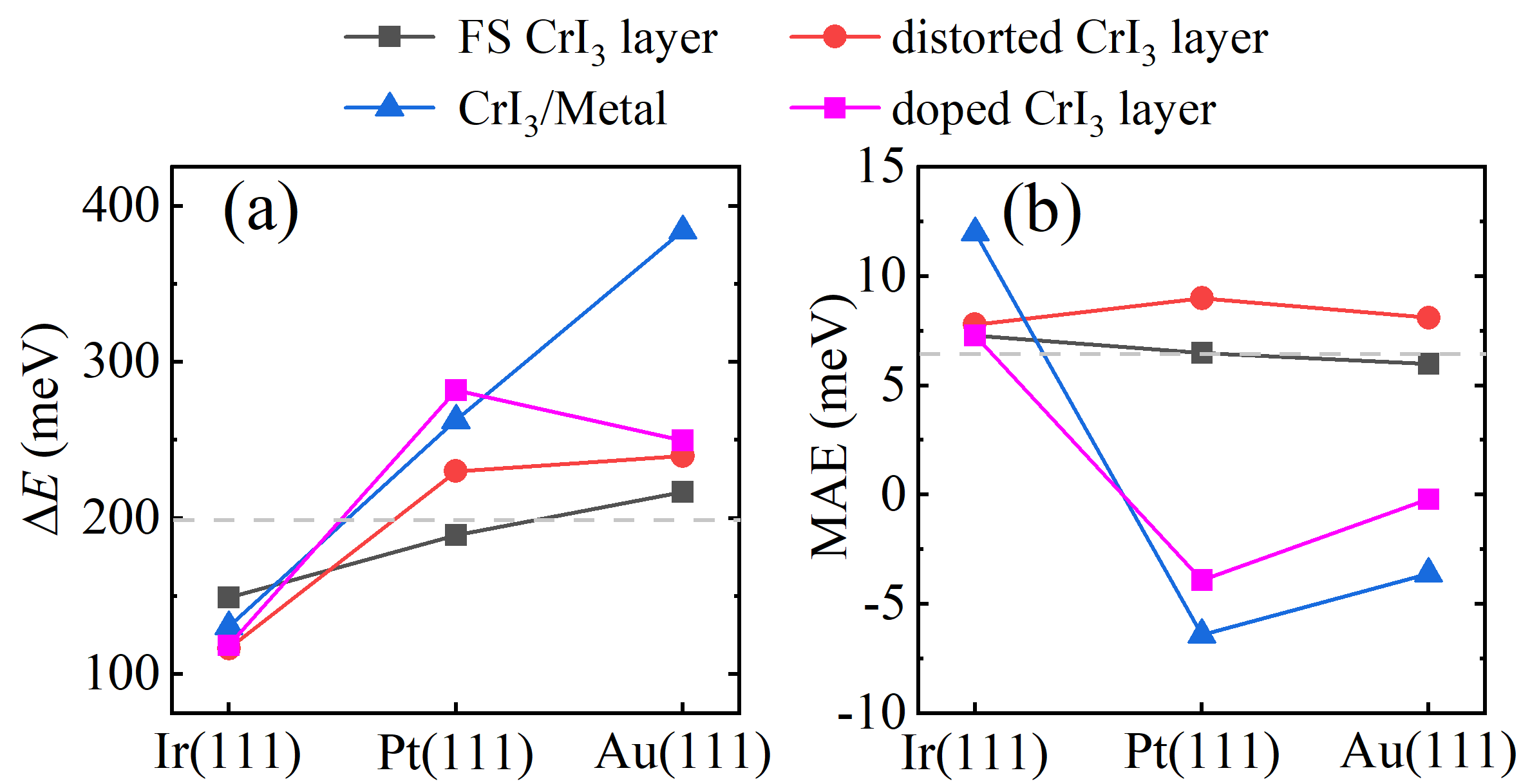}
\caption{The calculated exchange energy $\Delta$\emph{E} (a) and magnetic anisotropy energy MAE (b) for FS CrI$_3$ layer (blank), CrI$_3$ layer taken from CrI$_3$/Metal (red), CrI$_3$/Metal (blue) and charge doped CrI$_3$ layer (pink). The horizontal dash line represent the value of pristine CrI$_3$ layer.}
\label{Fig3}
\end{center}
\end{figure}

The strength of Heisenberg exchange interaction can be evaluated by the exchange energy, namely the total energy difference between N\'{e}el antiferromagnetic (AFM) and FM states, $\Delta$\emph{E}=\emph{E}$_{AFM}$-\emph{E}$_{FM}$. Magnetic anisotropy energy MAE is calculated by MAE=\emph{E}$_x$-\emph{E}$_z$, in which the \emph{E}$_z$ (\emph{E}$_x$) is total energy of the systems with out-of-plane (in-plane) magnetization in FM state. The results are plotted by red in Fig.~\ref{Fig3}. We find the $\Delta$\emph{E} increase by 32\% and 93\% in CrI$_3$/Pt(111) and CrI$_3$/Au(111), respectively, while CrI$_3$/Ir(111) has a small $\Delta$\emph{E} decreased by 34\%. Although the negative effect on $\Delta$\emph{E}, Ir(111) enhance the out-of-plane MAE from 6.4 meV for pristine CrI$_3$ monolayer to 12.0 meV. However, Pt(111) and Au(111) flip magnetism direction to in-plane.

Interfaces can result in several effects, including strain, interlayer hybridization and charge transfer. Understanding how is the magnetism influenced by these interfacial effects is important to the design of magnetic heterostructures. To achieve this target, we further perform magnetism calculations to the freestanding (FS) CrI$_3$ layer with the lattice constants that match to substrates, distorted CrI$_3$ layer taken from the relaxed heterostructures, and the distorted CrI$_3$ layer doped with charge as the charge transfer in the corresponding CrI$_3$/Metal. The results are shown by different colour in Fig.~\ref{Fig3}. For FS CrI$_3$ layer, the magnetism is modulated by lattice constant (strain). We can see that the $\Delta$\emph{E} increase with lattice constant, but a contrary tendency to MAE. This is consistent with the previous studies about the tuning of magnetism of CrI$_3$ layer by strain \cite{PRB_strain,wu2019strain}. The distorted CrI$_3$ layer further suffers structural distortions resulted by interlayer interactions. Interestingly, MAE is always enhanced by structural deformations, owning to the reduced crystal symmetry \cite{rau2014reaching}. The $\Delta$\emph{E} of CrI$_3$ layer taken from CrI$_3$/Pt(111) and CrI$_3$/Au(111) are larger than the corresponding FS CrI$_3$ layer, while a reduction of $\Delta$\emph{E} is caused by the crystal deformation in CrI$_3$/Ir(111). We can understand this discrepancy from a geometrical perspective. Upon structural relaxation, the change of upper Cr-I-Cr bond angle is negligible, but the bottom one becomes small for all the systems. Nevertheless, in CrI$_3$/Pt(111) (CrI$_3$/Au(111)), the bond angle change from 93.4$^{\circ}$ (97.0$^{\circ}$) to 89.2$^{\circ}$ (94.2$^{\circ}$), respectively, which are more close to the ferromagnetic criterion of 90$^{\circ}$ from GKA rule for superexchange \cite{GKA}. Conversely, the bond angle decreases from 91.4$^{\circ}$ to 84.6$^{\circ}$ by Ir(111), deviating largely from the ferromagnetic criterion. As a result, the ferromagnetism of Cr-I-Cr path is strengthen by Pt(111) and Au(111), while weaken by Ir(111).

\begin{figure*}
\begin{center}
\includegraphics[width=0.9\textwidth]{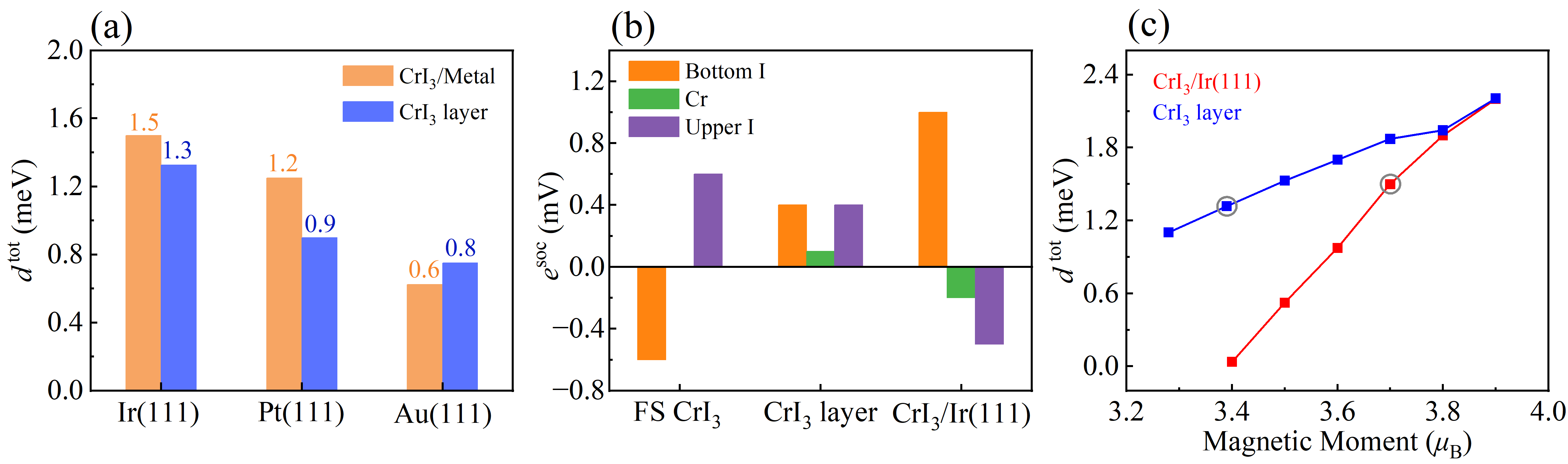}
\caption{(a) Total DMI of CrI$_3$/Metal and CrI$_3$ layer. (b) Atomic contribution to DMI. (c) DMI under different magnetic moments for CrI$_3$/Ir(111) and CrI$_3$ layer. The circled dots are the most energetically stable magnetic moments.}
\label{Fig4}
\end{center}
\end{figure*}

Then we further dope the distorted CrI$_3$ layer by the corresponding interlayer charge transfer. Because the charge transfer in CrI$_3$/Ir(111) is very small, the doping effect on $\Delta$\emph{E} and MAE is tiny. Meanwhile, the ferromagnetism is enhanced by both hole and electron doping, the same as previous studies \cite{kim2019exploitable,CrI3_GaN}. Interestingly, the $\Delta$\emph{E} of charge doped CrI$_3$ layer is much smaller than CrI$_3$/Au(111). Therefore, though the relative weak interlayer hybridization in CrI$_3$/Au(111), it plays dominant role to enhance the Heisenberg exchange interaction of CrI$_3$. By contrast, the $\Delta$\emph{E} of charge doped CrI$_3$ are close to CrI$_3$/Ir(111) and CrI$_3$/Pt(111), thus the strong interlayer hybridization between CrI$_3$ with Ir(111) and Pt(111) influence the Heisenberg exchange interaction slightly. On the other hand, the magnetism direction is flipped from out-of-plane to in-plane by both electron and hole doping. In addition, interlayer hybridization in CrI$_3$/Pt(111) and CrI$_3$/Au(111) can further stable the in-plane magnetism. Surprisingly, the magnetism direction maintains out-of-plane in CrI$_3$/Ir(111), and the MAE even increases by 88\% to a value of 12.0 meV. The out-of-plane anisotropy is crucial to the application of low dimensional magnetism ~\cite{fert2017magnetic}. Hence, CrI$_3$/Ir(111) is a potential candidate to fabrication of information device.

Another important effect in interface is the inversion symmetry breaking. In magnetic systems, combining with large enough SOC, Dzyaloshinskii-Moriya interaction (DMI) may arise. DMI among magnetic pair takes the form as \cite{pr1960}
\begin{equation}
E_{DM}=-\mathbf{D}_{ij}\cdot(\mathbf{S_i}\times\mathbf{S_j})
\end{equation}
which is antisymmetric and breaks chiral symmetry, thus the spin waves with opposite chirality are no long degenerate. On this account, we construct two short period spin waves with clockwise (cw) and anticlockwise (acw) chirality by the constrained magnetism as implemented in VASP \cite{PRL_Yang}, and calculate the corresponding total energy \emph{E}$_{cw}$ and \emph{E}$_{acw}$, then the total DMI energy can be expressed by
\begin{equation}
d^{tot}=(E_{acw}-E_{cw})/n
\end{equation}
where \emph{n}=8 is number of Cr atom in CrI$_3$/Metal. As shown in Fig.~\ref{Fig4}, CrI$_3$/Ir(111) has the largest \emph{d}$^{tot}$ of 1.5 meV, and decreases monotonously to 0.6 meV for CrI$_3$/Au(111). The strengths are comparable to the DMI in Janus \cite{MXY} and FM/HM heterostructures, such as Co/Pt and Co/Ir \cite{PRL_Yang}. Considering the out-of-plane magnetic anisotropy and weak Heisenberg exchange interaction, topological magnetic structure Skyrmions is likely to arise in CrI$_3$/Ir(111). On the other hand, vortex is possible in CrI$_3$/Pt(111) and CrI$_3$/Au(111) due to the in-plane magnetism. In 3\emph{d}/5\emph{d} interfaces, a similar trend of DMI for different substrates is also observed \cite{PRL3d_5d}, and the magnitude of DMI for 3\emph{d} atom layer deposited on Au(111) is always the smallest for the fulfilling 5\emph{d} orbital and the subsequent weak 3\emph{d}-5\emph{d} orbital hybridization \cite{PRL3d_5d,dmi_prl2018}. However, as discussed above, in the vdW interfaces Cr atoms do not interact with substrates directly, the hybridization between Cr 3\emph{d} and substrate 5\emph{d} orbital should not the key factor to generate DMI. To confirm this assumption, we calculate the DMI of distorted CrI$_3$ layer. The results show that CrI$_3$ layer also possesses sizable DMI that close to the interfaces, and the value is even greater CrI$_3$/Au(111). It should be noted that the MM of Cr atom in CrI$_3$/Metal is larger than CrI$_3$ layer. Consequently, we can conclude the structural distortion in CrI$_3$ layer caused by substrates take the major responsibility for the DMI in CrI$_3$/Metal. This is different from 3\emph{d}/5\emph{d} interfaces, in which the 3\emph{d}-5\emph{d} orbital hybridization is the crucial mechanism \cite{PRL_Yang,PRL3d_5d,PRL_5d}. The results manifest that we can induce DMI in 2D magnetic materials by choosing substrates with appropriate activity which can result in structural distortion, instead of the substrate with large SOC to provide spin-orbit center.

The DMI is rooted in the variation of atomic SOC energy at different spin configurations. To clarify atomic contribution to DMI, taking CrI$_3$/Ir(111) as an illustrating example, we calculate the atomic resolved SOC energies difference \emph{e}$^{soc}$ for opposite chirality spin waves, and the results are depicted in Fig.~\ref{Fig4}(b). In FS CrI$_3$, the \emph{e}$^{soc}$ of Cr atom is zero, and upper and bottom I atom layers have same absolute values but opposite signs. This result is expectant because FS CrI$_3$ is inversion symmetric and the DMI must be zero. As to distorted CrI$_3$ layer taken from relaxed CrI$_3$/Ir(111), the results show crystal distortion not only alters the magnitudes of \emph{e}$^{soc}$, but also reverses the sign of bottom I atom. In addition, the \emph{e}$^{soc}$ for Cr atom layer is no long zero with the magnitude about 0.2 meV, because the octahedral ligand field of Cr atom is distorted and the orbital moment is no long totally quenched. Consequently, a non-zero DMI is generated and contributed mainly by heavy I atoms. This is similar to the situation in Janus \cite{MXY,Janus_1}. Interestingly, once CrI$_3$ layer hybridizes with substrate, despite \emph{d}$^{tot}$ of CrI$_3$ layer and CrI$_3$/Ir(111) are close, \emph{e}$^{soc}$ alters dramatically. The \emph{e}$^{soc}$ of bottom I atom layer is more than doubled, while Cr and upper I atom change to negative sign. In addition, we should point out that the \emph{e}$^{soc}$ of substrates in all the systems is negligible. The heavy atoms with strong SOC is crucial to DMI since they act as spin-orbital scattering \cite{PRL1980}. However, in CrI$_3$/Metal, the Cr atom hybridize very weakly with substrates, so the superexchange path Cr-5\emph{d} atom-Cr does not develop. As a consequence, the contribution of substrates to DMI is tiny, even though strong SOC in substrates.

\begin{figure}
\begin{center}
\includegraphics[width=0.35\textwidth]{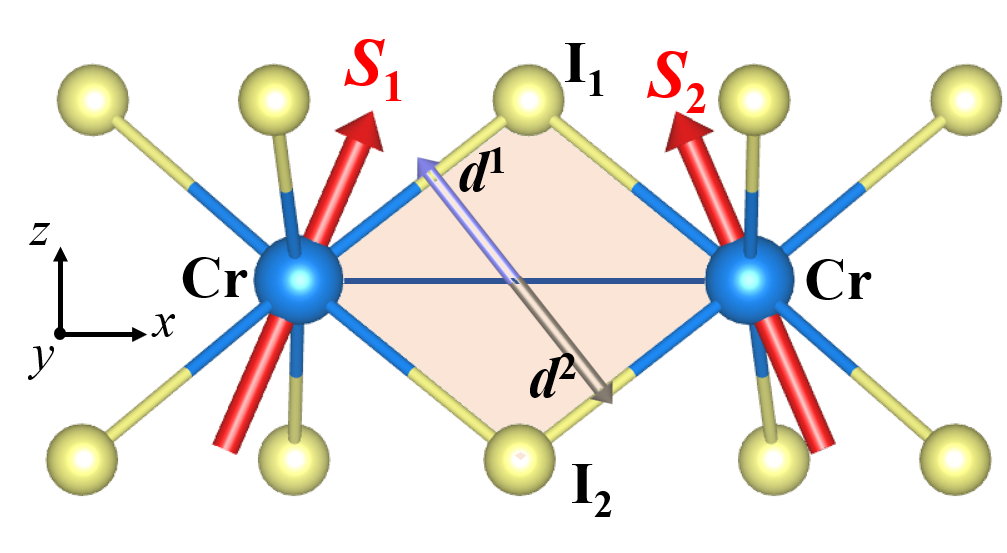}
\caption{Schematic show of DMI in CrI$_3$ layer.}
\label{Fig5}
\end{center}
\end{figure}

We further illustrate the DMI in Fig.~\ref{Fig5}. Among Cr pair where are two superexchange paths Cr-I$_1$(I$_2$)-Cr, and each path will generate DMI \emph{d}$^1$(\emph{d}$^2$). According to Moriya's rule, the direction of DMI is perpendicular the plane determined by the three atoms in path \cite{pr1960}. In FS CrI$_3$, the two paths are equal, so the \emph{d}$^1$ and \emph{d}$^2$ are opposite with the same magnitude, resulting in zero net DMI. However, in interfaces, because the structural distortion and interlayer hybridization, the two paths are no long equal and \emph{d}$^1$ can not cancel with \emph{d}$^2$. We note that net DMI may be tilted and has in-plane and out-of-plane components. Nevertheless, the average contribution of out-of-plane component to total energy is zero, and the above calculation is the in-plane component ~\cite{MXY}.

As stated above, the MMs of Cr atom in CrI$_3$ layer and CrI$_3$/Metal are different. To illustrate the dependence of DMI on MM, we derive the DMI as a function of the MM in Cr atom by constraining its magnitude for CrI$_3$/Ir(111), and the results are show in Fig.~\ref{Fig4}(c). Positive linear relationships are found for both CrI$_3$ layer and CrI$_3$/Ir(111), but the slope of CrI$_3$/Ir(111) is steeper than CrI$_3$ layer. This is different from Co/Pt interface, wherein DMI decreases even changes the sign along with the increasing of Co magnetic moment \cite{dmi2017PRB}. As expressed in eq. (3), when the MMs are the same, \emph{d}$^{tot}$ of CrI$_3$ layer is always greater than CrI$_3$/Ir(111). If $\mathbf{D}_{ij}$ is a constant, the value of energy is in direct proportion to the square of MM (spin). However, the linear relationships indicate the magnitude of $\mathbf{D}_{ij}$ is inversely proportional to the square root of MM.

\section{\label{sec:level1}Conclusions}
In summary, we have predicted large DMI in vdW magnetic heterostructures CrI$_3$/Metal. Our calculations and analysis indicate the underlaying mechanisms are the large SOC of I atom and structural distortion in CrI$_3$ layer, which are different to the traditional magnetic heterostructures where the DMI is generated by interlayer hybridization and the large SOC of substrates. In addition, both the Heisenberg exchange interaction and magnetic anisotropy energy can be tuned dramatically by substrates. This work will be helpful to the research of CrI$_3$ and the design of information device, since it provides an experimentally accessible way to induce DMI in vdW magnetic materials.

\emph{Note added.} During the preparation of our manuscript, Zhang et al.\cite{PRB2022} report the DMI in CrI$_3$ on Au and Cu substrates, but they attribute the mechanism to interlayer charge transfer and chemical interactions.

\begin{acknowledgments}
This work was supported by the National Natural Science Foundation of China (Grant No. 11874092 and 12174100), the Fok Ying-Tong Education Foundation, China (Grant No. 161005), and the Natural Science Foundation for Distinguished Young Scholars of Hunan Province (Grant No. 2021JJ10039).
\end{acknowledgments}

\bibliography{ref}

\begin{thebibliography}{40}%
\makeatletter
\providecommand \@ifxundefined [1]{%
 \@ifx{#1\undefined}
}%
\providecommand \@ifnum [1]{%
 \ifnum #1\expandafter \@firstoftwo
 \else \expandafter \@secondoftwo
 \fi
}%
\providecommand \@ifx [1]{%
 \ifx #1\expandafter \@firstoftwo
 \else \expandafter \@secondoftwo
 \fi
}%
\providecommand \natexlab [1]{#1}%
\providecommand \enquote  [1]{``#1''}%
\providecommand \bibnamefont  [1]{#1}%
\providecommand \bibfnamefont [1]{#1}%
\providecommand \citenamefont [1]{#1}%
\providecommand \href@noop [0]{\@secondoftwo}%
\providecommand \href [0]{\begingroup \@sanitize@url \@href}%
\providecommand \@href[1]{\@@startlink{#1}\@@href}%
\providecommand \@@href[1]{\endgroup#1\@@endlink}%
\providecommand \@sanitize@url [0]{\catcode `\\12\catcode `\$12\catcode
  `\&12\catcode `\#12\catcode `\^12\catcode `\_12\catcode `\%12\relax}%
\providecommand \@@startlink[1]{}%
\providecommand \@@endlink[0]{}%
\providecommand \url  [0]{\begingroup\@sanitize@url \@url }%
\providecommand \@url [1]{\endgroup\@href {#1}{\urlprefix }}%
\providecommand \urlprefix  [0]{URL }%
\providecommand \Eprint [0]{\href }%
\providecommand \doibase [0]{https://doi.org/}%
\providecommand \selectlanguage [0]{\@gobble}%
\providecommand \bibinfo  [0]{\@secondoftwo}%
\providecommand \bibfield  [0]{\@secondoftwo}%
\providecommand \translation [1]{[#1]}%
\providecommand \BibitemOpen [0]{}%
\providecommand \bibitemStop [0]{}%
\providecommand \bibitemNoStop [0]{.\EOS\space}%
\providecommand \EOS [0]{\spacefactor3000\relax}%
\providecommand \BibitemShut  [1]{\csname bibitem#1\endcsname}%
\let\auto@bib@innerbib\@empty
\bibitem [{\citenamefont {Kurebayashi}\ \emph {et~al.}(2022)\citenamefont
  {Kurebayashi}, \citenamefont {Garcia}, \citenamefont {Khan}, \citenamefont
  {Sinova},\ and\ \citenamefont {Roche}}]{2022magnetism}%
  \BibitemOpen
  \bibfield  {author} {\bibinfo {author} {\bibfnamefont {H.}~\bibnamefont
  {Kurebayashi}}, \bibinfo {author} {\bibfnamefont {J.~H.}\ \bibnamefont
  {Garcia}}, \bibinfo {author} {\bibfnamefont {S.}~\bibnamefont {Khan}},
  \bibinfo {author} {\bibfnamefont {J.}~\bibnamefont {Sinova}},\ and\ \bibinfo
  {author} {\bibfnamefont {S.}~\bibnamefont {Roche}},\ }\bibfield  {title}
  {\bibinfo {title} {Magnetism, symmetry and spin transport in van der waals
  layered systems},\ }\href@noop {} {\bibfield  {journal} {\bibinfo  {journal}
  {Nat. Rev. Phys.}\ }\textbf {\bibinfo {volume} {4}},\ \bibinfo {pages} {150}
  (\bibinfo {year} {2022})}\BibitemShut {NoStop}%
\bibitem [{\citenamefont {Khomskii}(2014)}]{TMC}%
  \BibitemOpen
  \bibfield  {author} {\bibinfo {author} {\bibfnamefont {D.}~\bibnamefont
  {Khomskii}},\ }\href@noop {} {\emph {\bibinfo {title} {Transition metal
  compounds}}}\ (\bibinfo  {publisher} {Cambridge University Press},\ \bibinfo
  {year} {2014})\BibitemShut {NoStop}%
\bibitem [{\citenamefont {Hellman}\ \emph {et~al.}(2017)\citenamefont
  {Hellman}, \citenamefont {Hoffmann}, \citenamefont {Tserkovnyak},
  \citenamefont {Beach}, \citenamefont {Fullerton}, \citenamefont {Leighton},
  \citenamefont {MacDonald}, \citenamefont {Ralph}, \citenamefont {Arena},
  \citenamefont {D\"urr}, \citenamefont {Fischer}, \citenamefont {Grollier},
  \citenamefont {Heremans}, \citenamefont {Jungwirth}, \citenamefont {Kimel},
  \citenamefont {Koopmans}, \citenamefont {Krivorotov}, \citenamefont {May},
  \citenamefont {Petford-Long}, \citenamefont {Rondinelli}, \citenamefont
  {Samarth}, \citenamefont {Schuller}, \citenamefont {Slavin}, \citenamefont
  {Stiles}, \citenamefont {Tchernyshyov}, \citenamefont {Thiaville},\ and\
  \citenamefont {Zink}}]{RMP2016}%
  \BibitemOpen
  \bibfield  {author} {\bibinfo {author} {\bibfnamefont {F.}~\bibnamefont
  {Hellman}}, \bibinfo {author} {\bibfnamefont {A.}~\bibnamefont {Hoffmann}},
  \bibinfo {author} {\bibfnamefont {Y.}~\bibnamefont {Tserkovnyak}}, \bibinfo
  {author} {\bibfnamefont {G.~S.~D.}\ \bibnamefont {Beach}}, \bibinfo {author}
  {\bibfnamefont {E.~E.}\ \bibnamefont {Fullerton}}, \bibinfo {author}
  {\bibfnamefont {C.}~\bibnamefont {Leighton}}, \bibinfo {author}
  {\bibfnamefont {A.~H.}\ \bibnamefont {MacDonald}}, \bibinfo {author}
  {\bibfnamefont {D.~C.}\ \bibnamefont {Ralph}}, \bibinfo {author}
  {\bibfnamefont {D.~A.}\ \bibnamefont {Arena}}, \bibinfo {author}
  {\bibfnamefont {H.~A.}\ \bibnamefont {D\"urr}}, \bibinfo {author}
  {\bibfnamefont {P.}~\bibnamefont {Fischer}}, \bibinfo {author} {\bibfnamefont
  {J.}~\bibnamefont {Grollier}}, \bibinfo {author} {\bibfnamefont {J.~P.}\
  \bibnamefont {Heremans}}, \bibinfo {author} {\bibfnamefont {T.}~\bibnamefont
  {Jungwirth}}, \bibinfo {author} {\bibfnamefont {A.~V.}\ \bibnamefont
  {Kimel}}, \bibinfo {author} {\bibfnamefont {B.}~\bibnamefont {Koopmans}},
  \bibinfo {author} {\bibfnamefont {I.~N.}\ \bibnamefont {Krivorotov}},
  \bibinfo {author} {\bibfnamefont {S.~J.}\ \bibnamefont {May}}, \bibinfo
  {author} {\bibfnamefont {A.~K.}\ \bibnamefont {Petford-Long}}, \bibinfo
  {author} {\bibfnamefont {J.~M.}\ \bibnamefont {Rondinelli}}, \bibinfo
  {author} {\bibfnamefont {N.}~\bibnamefont {Samarth}}, \bibinfo {author}
  {\bibfnamefont {I.~K.}\ \bibnamefont {Schuller}}, \bibinfo {author}
  {\bibfnamefont {A.~N.}\ \bibnamefont {Slavin}}, \bibinfo {author}
  {\bibfnamefont {M.~D.}\ \bibnamefont {Stiles}}, \bibinfo {author}
  {\bibfnamefont {O.}~\bibnamefont {Tchernyshyov}}, \bibinfo {author}
  {\bibfnamefont {A.}~\bibnamefont {Thiaville}},\ and\ \bibinfo {author}
  {\bibfnamefont {B.~L.}\ \bibnamefont {Zink}},\ }\bibfield  {title} {\bibinfo
  {title} {Interface-induced phenomena in magnetism},\ }\href
  {https://doi.org/10.1103/RevModPhys.89.025006} {\bibfield  {journal}
  {\bibinfo  {journal} {Rev. Mod. Phys.}\ }\textbf {\bibinfo {volume} {89}},\
  \bibinfo {pages} {025006} (\bibinfo {year} {2017})}\BibitemShut {NoStop}%
\bibitem [{\citenamefont {Heinze}\ \emph {et~al.}(2000)\citenamefont {Heinze},
  \citenamefont {Bode}, \citenamefont {Kubetzka}, \citenamefont {Pietzsch},
  \citenamefont {Nie}, \citenamefont {Blugel},\ and\ \citenamefont
  {Wiesendanger}}]{heinze2000real}%
  \BibitemOpen
  \bibfield  {author} {\bibinfo {author} {\bibfnamefont {S.}~\bibnamefont
  {Heinze}}, \bibinfo {author} {\bibfnamefont {M.}~\bibnamefont {Bode}},
  \bibinfo {author} {\bibfnamefont {A.}~\bibnamefont {Kubetzka}}, \bibinfo
  {author} {\bibfnamefont {O.}~\bibnamefont {Pietzsch}}, \bibinfo {author}
  {\bibfnamefont {X.}~\bibnamefont {Nie}}, \bibinfo {author} {\bibfnamefont
  {S.}~\bibnamefont {Blugel}},\ and\ \bibinfo {author} {\bibfnamefont
  {R.}~\bibnamefont {Wiesendanger}},\ }\bibfield  {title} {\bibinfo {title}
  {Real-space imaging of two-dimensional antiferromagnetism on the atomic
  scale},\ }\href@noop {} {\bibfield  {journal} {\bibinfo  {journal} {Science}\
  }\textbf {\bibinfo {volume} {288}},\ \bibinfo {pages} {1805} (\bibinfo {year}
  {2000})}\BibitemShut {NoStop}%
\bibitem [{\citenamefont {Fert}\ \emph {et~al.}(2017)\citenamefont {Fert},
  \citenamefont {Reyren},\ and\ \citenamefont {Cros}}]{fert2017magnetic}%
  \BibitemOpen
  \bibfield  {author} {\bibinfo {author} {\bibfnamefont {A.}~\bibnamefont
  {Fert}}, \bibinfo {author} {\bibfnamefont {N.}~\bibnamefont {Reyren}},\ and\
  \bibinfo {author} {\bibfnamefont {V.}~\bibnamefont {Cros}},\ }\bibfield
  {title} {\bibinfo {title} {Magnetic skyrmions: advances in physics and
  potential applications},\ }\href@noop {} {\bibfield  {journal} {\bibinfo
  {journal} {Nat. Rev. Mater.}\ }\textbf {\bibinfo {volume} {2}},\ \bibinfo
  {pages} {1} (\bibinfo {year} {2017})}\BibitemShut {NoStop}%
\bibitem [{\citenamefont {Heinze}\ \emph {et~al.}(2011)\citenamefont {Heinze},
  \citenamefont {Von~Bergmann}, \citenamefont {Menzel}, \citenamefont {Brede},
  \citenamefont {Kubetzka}, \citenamefont {Wiesendanger}, \citenamefont
  {Bihlmayer},\ and\ \citenamefont {Bl{\"u}gel}}]{heinze2011spontaneous}%
  \BibitemOpen
  \bibfield  {author} {\bibinfo {author} {\bibfnamefont {S.}~\bibnamefont
  {Heinze}}, \bibinfo {author} {\bibfnamefont {K.}~\bibnamefont
  {Von~Bergmann}}, \bibinfo {author} {\bibfnamefont {M.}~\bibnamefont
  {Menzel}}, \bibinfo {author} {\bibfnamefont {J.}~\bibnamefont {Brede}},
  \bibinfo {author} {\bibfnamefont {A.}~\bibnamefont {Kubetzka}}, \bibinfo
  {author} {\bibfnamefont {R.}~\bibnamefont {Wiesendanger}}, \bibinfo {author}
  {\bibfnamefont {G.}~\bibnamefont {Bihlmayer}},\ and\ \bibinfo {author}
  {\bibfnamefont {S.}~\bibnamefont {Bl{\"u}gel}},\ }\bibfield  {title}
  {\bibinfo {title} {Spontaneous atomic-scale magnetic skyrmion lattice in two
  dimensions},\ }\href@noop {} {\bibfield  {journal} {\bibinfo  {journal} {Nat.
  Phys.}\ }\textbf {\bibinfo {volume} {7}},\ \bibinfo {pages} {713} (\bibinfo
  {year} {2011})}\BibitemShut {NoStop}%
\bibitem [{\citenamefont {Belabbes}\ \emph {et~al.}(2016)\citenamefont
  {Belabbes}, \citenamefont {Bihlmayer}, \citenamefont {Bechstedt},
  \citenamefont {Bl\"ugel},\ and\ \citenamefont {Manchon}}]{PRL3d_5d}%
  \BibitemOpen
  \bibfield  {author} {\bibinfo {author} {\bibfnamefont {A.}~\bibnamefont
  {Belabbes}}, \bibinfo {author} {\bibfnamefont {G.}~\bibnamefont {Bihlmayer}},
  \bibinfo {author} {\bibfnamefont {F.}~\bibnamefont {Bechstedt}}, \bibinfo
  {author} {\bibfnamefont {S.}~\bibnamefont {Bl\"ugel}},\ and\ \bibinfo
  {author} {\bibfnamefont {A.}~\bibnamefont {Manchon}},\ }\bibfield  {title}
  {\bibinfo {title} {Hund's rule-driven {Dzyaloshinskii-Moriya} interaction at
  $3d\text{\ensuremath{-}}5d$ interfaces},\ }\href
  {https://doi.org/10.1103/PhysRevLett.117.247202} {\bibfield  {journal}
  {\bibinfo  {journal} {Phys. Rev. Lett.}\ }\textbf {\bibinfo {volume} {117}},\
  \bibinfo {pages} {247202} (\bibinfo {year} {2016})}\BibitemShut {NoStop}%
\bibitem [{\citenamefont {Yang}\ \emph {et~al.}(2015)\citenamefont {Yang},
  \citenamefont {Thiaville}, \citenamefont {Rohart}, \citenamefont {Fert},\
  and\ \citenamefont {Chshiev}}]{PRL_Yang}%
  \BibitemOpen
  \bibfield  {author} {\bibinfo {author} {\bibfnamefont {H.}~\bibnamefont
  {Yang}}, \bibinfo {author} {\bibfnamefont {A.}~\bibnamefont {Thiaville}},
  \bibinfo {author} {\bibfnamefont {S.}~\bibnamefont {Rohart}}, \bibinfo
  {author} {\bibfnamefont {A.}~\bibnamefont {Fert}},\ and\ \bibinfo {author}
  {\bibfnamefont {M.}~\bibnamefont {Chshiev}},\ }\bibfield  {title} {\bibinfo
  {title} {Anatomy of {Dzyaloshinskii-Moriya} interaction at
  $\mathrm{Co}/\mathrm{Pt}$ interfaces},\ }\href
  {https://doi.org/10.1103/PhysRevLett.115.267210} {\bibfield  {journal}
  {\bibinfo  {journal} {Phys. Rev. Lett.}\ }\textbf {\bibinfo {volume} {115}},\
  \bibinfo {pages} {267210} (\bibinfo {year} {2015})}\BibitemShut {NoStop}%
\bibitem [{\citenamefont {Huang}\ \emph {et~al.}(2017)\citenamefont {Huang},
  \citenamefont {Clark}, \citenamefont {Navarro-Moratalla}, \citenamefont
  {Klein}, \citenamefont {Cheng}, \citenamefont {Seyler}, \citenamefont
  {Zhong}, \citenamefont {Schmidgall}, \citenamefont {McGuire}, \citenamefont
  {Cobden} \emph {et~al.}}]{huang2017layer}%
  \BibitemOpen
  \bibfield  {author} {\bibinfo {author} {\bibfnamefont {B.}~\bibnamefont
  {Huang}}, \bibinfo {author} {\bibfnamefont {G.}~\bibnamefont {Clark}},
  \bibinfo {author} {\bibfnamefont {E.}~\bibnamefont {Navarro-Moratalla}},
  \bibinfo {author} {\bibfnamefont {D.~R.}\ \bibnamefont {Klein}}, \bibinfo
  {author} {\bibfnamefont {R.}~\bibnamefont {Cheng}}, \bibinfo {author}
  {\bibfnamefont {K.~L.}\ \bibnamefont {Seyler}}, \bibinfo {author}
  {\bibfnamefont {D.}~\bibnamefont {Zhong}}, \bibinfo {author} {\bibfnamefont
  {E.}~\bibnamefont {Schmidgall}}, \bibinfo {author} {\bibfnamefont {M.~A.}\
  \bibnamefont {McGuire}}, \bibinfo {author} {\bibfnamefont {D.~H.}\
  \bibnamefont {Cobden}}, \emph {et~al.},\ }\bibfield  {title} {\bibinfo
  {title} {Layer-dependent ferromagnetism in a van der {W}aals crystal down to
  the monolayer limit},\ }\href@noop {} {\bibfield  {journal} {\bibinfo
  {journal} {Nature}\ }\textbf {\bibinfo {volume} {546}},\ \bibinfo {pages}
  {270} (\bibinfo {year} {2017})}\BibitemShut {NoStop}%
\bibitem [{\citenamefont {Gong}\ \emph {et~al.}(2017)\citenamefont {Gong},
  \citenamefont {Li}, \citenamefont {Li}, \citenamefont {Ji}, \citenamefont
  {Stern}, \citenamefont {Xia}, \citenamefont {Cao}, \citenamefont {Bao},
  \citenamefont {Wang}, \citenamefont {Wang} \emph
  {et~al.}}]{gong2017discovery}%
  \BibitemOpen
  \bibfield  {author} {\bibinfo {author} {\bibfnamefont {C.}~\bibnamefont
  {Gong}}, \bibinfo {author} {\bibfnamefont {L.}~\bibnamefont {Li}}, \bibinfo
  {author} {\bibfnamefont {Z.}~\bibnamefont {Li}}, \bibinfo {author}
  {\bibfnamefont {H.}~\bibnamefont {Ji}}, \bibinfo {author} {\bibfnamefont
  {A.}~\bibnamefont {Stern}}, \bibinfo {author} {\bibfnamefont
  {Y.}~\bibnamefont {Xia}}, \bibinfo {author} {\bibfnamefont {T.}~\bibnamefont
  {Cao}}, \bibinfo {author} {\bibfnamefont {W.}~\bibnamefont {Bao}}, \bibinfo
  {author} {\bibfnamefont {C.}~\bibnamefont {Wang}}, \bibinfo {author}
  {\bibfnamefont {Y.}~\bibnamefont {Wang}}, \emph {et~al.},\ }\bibfield
  {title} {\bibinfo {title} {Discovery of intrinsic ferromagnetism in
  two-dimensional van der waals crystals},\ }\href@noop {} {\bibfield
  {journal} {\bibinfo  {journal} {Nature}\ }\textbf {\bibinfo {volume} {546}},\
  \bibinfo {pages} {265} (\bibinfo {year} {2017})}\BibitemShut {NoStop}%
\bibitem [{\citenamefont {Zhang}\ \emph {et~al.}(2015)\citenamefont {Zhang},
  \citenamefont {Qu}, \citenamefont {Zhu},\ and\ \citenamefont
  {Lam}}]{jmcc2016}%
  \BibitemOpen
  \bibfield  {author} {\bibinfo {author} {\bibfnamefont {W.-B.}\ \bibnamefont
  {Zhang}}, \bibinfo {author} {\bibfnamefont {Q.}~\bibnamefont {Qu}}, \bibinfo
  {author} {\bibfnamefont {P.}~\bibnamefont {Zhu}},\ and\ \bibinfo {author}
  {\bibfnamefont {C.-H.}\ \bibnamefont {Lam}},\ }\bibfield  {title} {\bibinfo
  {title} {Robust intrinsic ferromagnetism and half semiconductivity in stable
  two-dimensional single-layer chromium trihalides},\ }\href@noop {} {\bibfield
   {journal} {\bibinfo  {journal} {J. Mater. Chem. C}\ }\textbf {\bibinfo
  {volume} {3}},\ \bibinfo {pages} {12457} (\bibinfo {year}
  {2015})}\BibitemShut {NoStop}%
\bibitem [{\citenamefont {Kim}\ \emph {et~al.}(2019{\natexlab{a}})\citenamefont
  {Kim}, \citenamefont {Kim}, \citenamefont {Ko}, \citenamefont {Seo},
  \citenamefont {Kim}, \citenamefont {Jang}, \citenamefont {Kim}, \citenamefont
  {Kim}, \citenamefont {Cheong},\ and\ \citenamefont {Park}}]{CrI3_MAE}%
  \BibitemOpen
  \bibfield  {author} {\bibinfo {author} {\bibfnamefont {D.-H.}\ \bibnamefont
  {Kim}}, \bibinfo {author} {\bibfnamefont {K.}~\bibnamefont {Kim}}, \bibinfo
  {author} {\bibfnamefont {K.-T.}\ \bibnamefont {Ko}}, \bibinfo {author}
  {\bibfnamefont {J.}~\bibnamefont {Seo}}, \bibinfo {author} {\bibfnamefont
  {J.~S.}\ \bibnamefont {Kim}}, \bibinfo {author} {\bibfnamefont {T.-H.}\
  \bibnamefont {Jang}}, \bibinfo {author} {\bibfnamefont {Y.}~\bibnamefont
  {Kim}}, \bibinfo {author} {\bibfnamefont {J.-Y.}\ \bibnamefont {Kim}},
  \bibinfo {author} {\bibfnamefont {S.-W.}\ \bibnamefont {Cheong}},\ and\
  \bibinfo {author} {\bibfnamefont {J.-H.}\ \bibnamefont {Park}},\ }\bibfield
  {title} {\bibinfo {title} {Giant magnetic anisotropy induced by ligand {$LS$}
  coupling in layered {Cr} compounds},\ }\href
  {https://doi.org/10.1103/PhysRevLett.122.207201} {\bibfield  {journal}
  {\bibinfo  {journal} {Phys. Rev. Lett.}\ }\textbf {\bibinfo {volume} {122}},\
  \bibinfo {pages} {207201} (\bibinfo {year} {2019}{\natexlab{a}})}\BibitemShut
  {NoStop}%
\bibitem [{\citenamefont {Liu}\ \emph {et~al.}(2018)\citenamefont {Liu},
  \citenamefont {Shi}, \citenamefont {Lu},\ and\ \citenamefont
  {Anantram}}]{PRB_Liu}%
  \BibitemOpen
  \bibfield  {author} {\bibinfo {author} {\bibfnamefont {J.}~\bibnamefont
  {Liu}}, \bibinfo {author} {\bibfnamefont {M.}~\bibnamefont {Shi}}, \bibinfo
  {author} {\bibfnamefont {J.}~\bibnamefont {Lu}},\ and\ \bibinfo {author}
  {\bibfnamefont {M.~P.}\ \bibnamefont {Anantram}},\ }\bibfield  {title}
  {\bibinfo {title} {Analysis of electrical-field-dependent
  dzyaloshinskii-moriya interaction and magnetocrystalline anisotropy in a
  two-dimensional ferromagnetic monolayer},\ }\href
  {https://doi.org/10.1103/PhysRevB.97.054416} {\bibfield  {journal} {\bibinfo
  {journal} {Phys. Rev. B}\ }\textbf {\bibinfo {volume} {97}},\ \bibinfo
  {pages} {054416} (\bibinfo {year} {2018})}\BibitemShut {NoStop}%
\bibitem [{\citenamefont {Behera}\ \emph {et~al.}(2019)\citenamefont {Behera},
  \citenamefont {Chowdhury},\ and\ \citenamefont {Das}}]{APL}%
  \BibitemOpen
  \bibfield  {author} {\bibinfo {author} {\bibfnamefont {A.~K.}\ \bibnamefont
  {Behera}}, \bibinfo {author} {\bibfnamefont {S.}~\bibnamefont {Chowdhury}},\
  and\ \bibinfo {author} {\bibfnamefont {S.~R.}\ \bibnamefont {Das}},\
  }\bibfield  {title} {\bibinfo {title} {Magnetic skyrmions in atomic thin
  {CrI$_3$} monolayer},\ }\href@noop {} {\bibfield  {journal} {\bibinfo
  {journal} {Appl. Phys. Lett.}\ }\textbf {\bibinfo {volume} {114}},\ \bibinfo
  {pages} {232402} (\bibinfo {year} {2019})}\BibitemShut {NoStop}%
\bibitem [{\citenamefont {Xu}\ \emph {et~al.}(2020)\citenamefont {Xu},
  \citenamefont {Feng}, \citenamefont {Prokhorenko}, \citenamefont {Nahas},
  \citenamefont {Xiang},\ and\ \citenamefont {Bellaiche}}]{CrIX3}%
  \BibitemOpen
  \bibfield  {author} {\bibinfo {author} {\bibfnamefont {C.}~\bibnamefont
  {Xu}}, \bibinfo {author} {\bibfnamefont {J.}~\bibnamefont {Feng}}, \bibinfo
  {author} {\bibfnamefont {S.}~\bibnamefont {Prokhorenko}}, \bibinfo {author}
  {\bibfnamefont {Y.}~\bibnamefont {Nahas}}, \bibinfo {author} {\bibfnamefont
  {H.}~\bibnamefont {Xiang}},\ and\ \bibinfo {author} {\bibfnamefont
  {L.}~\bibnamefont {Bellaiche}},\ }\bibfield  {title} {\bibinfo {title}
  {Topological spin texture in {J}anus monolayers of the chromium trihalides
  {Cr(I, X)$_3$}},\ }\href {https://doi.org/10.1103/PhysRevB.101.060404}
  {\bibfield  {journal} {\bibinfo  {journal} {Phys. Rev. B}\ }\textbf {\bibinfo
  {volume} {101}},\ \bibinfo {pages} {060404} (\bibinfo {year}
  {2020})}\BibitemShut {NoStop}%
\bibitem [{\citenamefont {Li}\ \emph {et~al.}(2020{\natexlab{a}})\citenamefont
  {Li}, \citenamefont {Wang}, \citenamefont {Zhang}, \citenamefont {Chen},
  \citenamefont {Guo}, \citenamefont {Ji},\ and\ \citenamefont
  {Zhong}}]{li2020single}%
  \BibitemOpen
  \bibfield  {author} {\bibinfo {author} {\bibfnamefont {P.}~\bibnamefont
  {Li}}, \bibinfo {author} {\bibfnamefont {C.}~\bibnamefont {Wang}}, \bibinfo
  {author} {\bibfnamefont {J.}~\bibnamefont {Zhang}}, \bibinfo {author}
  {\bibfnamefont {S.}~\bibnamefont {Chen}}, \bibinfo {author} {\bibfnamefont
  {D.}~\bibnamefont {Guo}}, \bibinfo {author} {\bibfnamefont {W.}~\bibnamefont
  {Ji}},\ and\ \bibinfo {author} {\bibfnamefont {D.}~\bibnamefont {Zhong}},\
  }\bibfield  {title} {\bibinfo {title} {Single-layer {CrI$_3$} grown by
  molecular beam epitaxy},\ }\href@noop {} {\bibfield  {journal} {\bibinfo
  {journal} {Sci. Bull.}\ }\textbf {\bibinfo {volume} {65}},\ \bibinfo {pages}
  {1064} (\bibinfo {year} {2020}{\natexlab{a}})}\BibitemShut {NoStop}%
\bibitem [{\citenamefont {Kresse}\ and\ \citenamefont {Hafner}(1993)}]{vasp1}%
  \BibitemOpen
  \bibfield  {author} {\bibinfo {author} {\bibfnamefont {G.}~\bibnamefont
  {Kresse}}\ and\ \bibinfo {author} {\bibfnamefont {J.}~\bibnamefont
  {Hafner}},\ }\bibfield  {title} {\bibinfo {title} {Ab initio molecular
  dynamics for liquid metals},\ }\href
  {https://doi.org/10.1103/PhysRevB.47.558} {\bibfield  {journal} {\bibinfo
  {journal} {Phys. Rev. B}\ }\textbf {\bibinfo {volume} {47}},\ \bibinfo
  {pages} {558} (\bibinfo {year} {1993})}\BibitemShut {NoStop}%
\bibitem [{\citenamefont {Kresse}\ and\ \citenamefont
  {Furthm\"uller}(1996)}]{vasp2}%
  \BibitemOpen
  \bibfield  {author} {\bibinfo {author} {\bibfnamefont {G.}~\bibnamefont
  {Kresse}}\ and\ \bibinfo {author} {\bibfnamefont {J.}~\bibnamefont
  {Furthm\"uller}},\ }\bibfield  {title} {\bibinfo {title} {Efficient iterative
  schemes for ab initio total-energy calculations using a plane-wave basis
  set},\ }\href@noop {} {\bibfield  {journal} {\bibinfo  {journal} {Phys. Rev.
  B}\ }\textbf {\bibinfo {volume} {54}},\ \bibinfo {pages} {11169} (\bibinfo
  {year} {1996})}\BibitemShut {NoStop}%
\bibitem [{\citenamefont {Kresse}\ and\ \citenamefont {Joubert}(1999)}]{paw}%
  \BibitemOpen
  \bibfield  {author} {\bibinfo {author} {\bibfnamefont {G.}~\bibnamefont
  {Kresse}}\ and\ \bibinfo {author} {\bibfnamefont {D.}~\bibnamefont
  {Joubert}},\ }\bibfield  {title} {\bibinfo {title} {From ultrasoft
  pseudopotentials to the projector augmented-wave method},\ }\href@noop {}
  {\bibfield  {journal} {\bibinfo  {journal} {Phys. Rev. B}\ }\textbf {\bibinfo
  {volume} {59}},\ \bibinfo {pages} {1758} (\bibinfo {year}
  {1999})}\BibitemShut {NoStop}%
\bibitem [{\citenamefont {Perdew}\ \emph {et~al.}(1996)\citenamefont {Perdew},
  \citenamefont {Burke},\ and\ \citenamefont {Ernzerhof}}]{gga}%
  \BibitemOpen
  \bibfield  {author} {\bibinfo {author} {\bibfnamefont {J.~P.}\ \bibnamefont
  {Perdew}}, \bibinfo {author} {\bibfnamefont {K.}~\bibnamefont {Burke}},\ and\
  \bibinfo {author} {\bibfnamefont {M.}~\bibnamefont {Ernzerhof}},\ }\bibfield
  {title} {\bibinfo {title} {Generalized gradient approximation made simple},\
  }\href@noop {} {\bibfield  {journal} {\bibinfo  {journal} {Phys. Rev. Lett.}\
  }\textbf {\bibinfo {volume} {77}},\ \bibinfo {pages} {3865} (\bibinfo {year}
  {1996})}\BibitemShut {NoStop}%
\bibitem [{\citenamefont {Dudarev}\ \emph {et~al.}(1998)\citenamefont
  {Dudarev}, \citenamefont {Botton}, \citenamefont {Savrasov}, \citenamefont
  {Humphreys},\ and\ \citenamefont {Sutton}}]{U}%
  \BibitemOpen
  \bibfield  {author} {\bibinfo {author} {\bibfnamefont {S.~L.}\ \bibnamefont
  {Dudarev}}, \bibinfo {author} {\bibfnamefont {G.~A.}\ \bibnamefont {Botton}},
  \bibinfo {author} {\bibfnamefont {S.~Y.}\ \bibnamefont {Savrasov}}, \bibinfo
  {author} {\bibfnamefont {C.~J.}\ \bibnamefont {Humphreys}},\ and\ \bibinfo
  {author} {\bibfnamefont {A.~P.}\ \bibnamefont {Sutton}},\ }\bibfield  {title}
  {\bibinfo {title} {Electron-energy-loss spectra and the structural stability
  of nickel oxide: {An LSDA+U study}},\ }\href
  {https://doi.org/10.1103/PhysRevB.57.1505} {\bibfield  {journal} {\bibinfo
  {journal} {Phys. Rev. B}\ }\textbf {\bibinfo {volume} {57}},\ \bibinfo
  {pages} {1505} (\bibinfo {year} {1998})}\BibitemShut {NoStop}%
\bibitem [{\citenamefont {Monkhorst}\ and\ \citenamefont
  {Pack}(1976)}]{kpoint}%
  \BibitemOpen
  \bibfield  {author} {\bibinfo {author} {\bibfnamefont {H.~J.}\ \bibnamefont
  {Monkhorst}}\ and\ \bibinfo {author} {\bibfnamefont {J.~D.}\ \bibnamefont
  {Pack}},\ }\bibfield  {title} {\bibinfo {title} {Special points for
  brillouin-zone integrations},\ }\href@noop {} {\bibfield  {journal} {\bibinfo
   {journal} {Phys. Rev. B}\ }\textbf {\bibinfo {volume} {13}},\ \bibinfo
  {pages} {5188} (\bibinfo {year} {1976})}\BibitemShut {NoStop}%
\bibitem [{\citenamefont {Grimme}(2006)}]{d2}%
  \BibitemOpen
  \bibfield  {author} {\bibinfo {author} {\bibfnamefont {S.}~\bibnamefont
  {Grimme}},\ }\bibfield  {title} {\bibinfo {title} {Semiempirical {GGA-type}
  density functional constructed with a long-range dispersion correction},\
  }\href@noop {} {\bibfield  {journal} {\bibinfo  {journal} {J. Comput. Chem.}\
  }\textbf {\bibinfo {volume} {27}},\ \bibinfo {pages} {1787} (\bibinfo {year}
  {2006})}\BibitemShut {NoStop}%
\bibitem [{\citenamefont {Li}\ \emph {et~al.}(2019)\citenamefont {Li},
  \citenamefont {Xu}, \citenamefont {Lai},\ and\ \citenamefont
  {Zhang}}]{li2019enhanced}%
  \BibitemOpen
  \bibfield  {author} {\bibinfo {author} {\bibfnamefont {H.}~\bibnamefont
  {Li}}, \bibinfo {author} {\bibfnamefont {Y.-K.}\ \bibnamefont {Xu}}, \bibinfo
  {author} {\bibfnamefont {K.}~\bibnamefont {Lai}},\ and\ \bibinfo {author}
  {\bibfnamefont {W.-B.}\ \bibnamefont {Zhang}},\ }\bibfield  {title} {\bibinfo
  {title} {The enhanced ferromagnetism of single-layer {CrX$_3$ (X=Br and I)}
  via van der waals engineering},\ }\href@noop {} {\bibfield  {journal}
  {\bibinfo  {journal} {Phys. Chem. Chem. Phys.}\ }\textbf {\bibinfo {volume}
  {21}},\ \bibinfo {pages} {11949} (\bibinfo {year} {2019})}\BibitemShut
  {NoStop}%
\bibitem [{\citenamefont {Li}\ \emph {et~al.}(2020{\natexlab{b}})\citenamefont
  {Li}, \citenamefont {Xu}, \citenamefont {Cheng}, \citenamefont {He},\ and\
  \citenamefont {Zhang}}]{li2020spin}%
  \BibitemOpen
  \bibfield  {author} {\bibinfo {author} {\bibfnamefont {H.}~\bibnamefont
  {Li}}, \bibinfo {author} {\bibfnamefont {Y.-K.}\ \bibnamefont {Xu}}, \bibinfo
  {author} {\bibfnamefont {Z.-P.}\ \bibnamefont {Cheng}}, \bibinfo {author}
  {\bibfnamefont {B.-G.}\ \bibnamefont {He}},\ and\ \bibinfo {author}
  {\bibfnamefont {W.-B.}\ \bibnamefont {Zhang}},\ }\bibfield  {title} {\bibinfo
  {title} {Spin-dependent schottky barriers and vacancy-induced spin-selective
  ohmic contacts in magnetic {vdW} heterostructures},\ }\href@noop {}
  {\bibfield  {journal} {\bibinfo  {journal} {Phys. Chem. Chem. Phys.}\
  }\textbf {\bibinfo {volume} {22}},\ \bibinfo {pages} {9460} (\bibinfo {year}
  {2020}{\natexlab{b}})}\BibitemShut {NoStop}%
\bibitem [{\citenamefont {Ye}\ \emph {et~al.}(2021)\citenamefont {Ye},
  \citenamefont {Wang}, \citenamefont {Bai}, \citenamefont {Zhang},
  \citenamefont {Wu}, \citenamefont {Zhang},\ and\ \citenamefont
  {Wang}}]{CrI3_GaN}%
  \BibitemOpen
  \bibfield  {author} {\bibinfo {author} {\bibfnamefont {H.}~\bibnamefont
  {Ye}}, \bibinfo {author} {\bibfnamefont {X.}~\bibnamefont {Wang}}, \bibinfo
  {author} {\bibfnamefont {D.}~\bibnamefont {Bai}}, \bibinfo {author}
  {\bibfnamefont {J.}~\bibnamefont {Zhang}}, \bibinfo {author} {\bibfnamefont
  {X.}~\bibnamefont {Wu}}, \bibinfo {author} {\bibfnamefont {G.~P.}\
  \bibnamefont {Zhang}},\ and\ \bibinfo {author} {\bibfnamefont
  {J.}~\bibnamefont {Wang}},\ }\bibfield  {title} {\bibinfo {title}
  {Significant enhancement of magnetic anisotropy and conductivity in
  {GaN/CrI$_3$} van der waals heterostructures via electrostatic doping},\
  }\href {https://doi.org/10.1103/PhysRevB.104.075433} {\bibfield  {journal}
  {\bibinfo  {journal} {Phys. Rev. B}\ }\textbf {\bibinfo {volume} {104}},\
  \bibinfo {pages} {075433} (\bibinfo {year} {2021})}\BibitemShut {NoStop}%
\bibitem [{\citenamefont {Henkelman}\ \emph {et~al.}(2006)\citenamefont
  {Henkelman}, \citenamefont {Arnaldsson},\ and\ \citenamefont
  {J{\'o}nsson}}]{bader}%
  \BibitemOpen
  \bibfield  {author} {\bibinfo {author} {\bibfnamefont {G.}~\bibnamefont
  {Henkelman}}, \bibinfo {author} {\bibfnamefont {A.}~\bibnamefont
  {Arnaldsson}},\ and\ \bibinfo {author} {\bibfnamefont {H.}~\bibnamefont
  {J{\'o}nsson}},\ }\bibfield  {title} {\bibinfo {title} {A fast and robust
  algorithm for bader decomposition of charge density},\ }\href@noop {}
  {\bibfield  {journal} {\bibinfo  {journal} {Comp. Mater. Sci.}\ }\textbf
  {\bibinfo {volume} {36}},\ \bibinfo {pages} {354} (\bibinfo {year}
  {2006})}\BibitemShut {NoStop}%
\bibitem [{\citenamefont {Webster}\ and\ \citenamefont
  {Yan}(2018)}]{PRB_strain}%
  \BibitemOpen
  \bibfield  {author} {\bibinfo {author} {\bibfnamefont {L.}~\bibnamefont
  {Webster}}\ and\ \bibinfo {author} {\bibfnamefont {J.-A.}\ \bibnamefont
  {Yan}},\ }\bibfield  {title} {\bibinfo {title} {Strain-tunable magnetic
  anisotropy in monolayer {CrCl$_3$}, {CrBr$_3$}, and {CrI$_3$}},\ }\href
  {https://doi.org/10.1103/PhysRevB.98.144411} {\bibfield  {journal} {\bibinfo
  {journal} {Phys. Rev. B}\ }\textbf {\bibinfo {volume} {98}},\ \bibinfo
  {pages} {144411} (\bibinfo {year} {2018})}\BibitemShut {NoStop}%
\bibitem [{\citenamefont {Wu}\ \emph {et~al.}(2019)\citenamefont {Wu},
  \citenamefont {Yu},\ and\ \citenamefont {Yuan}}]{wu2019strain}%
  \BibitemOpen
  \bibfield  {author} {\bibinfo {author} {\bibfnamefont {Z.}~\bibnamefont
  {Wu}}, \bibinfo {author} {\bibfnamefont {J.}~\bibnamefont {Yu}},\ and\
  \bibinfo {author} {\bibfnamefont {S.}~\bibnamefont {Yuan}},\ }\bibfield
  {title} {\bibinfo {title} {Strain-tunable magnetic and electronic properties
  of monolayer {CrI$_3$}},\ }\href@noop {} {\bibfield  {journal} {\bibinfo
  {journal} {Phys. Chem. Chem. Phys.}\ }\textbf {\bibinfo {volume} {21}},\
  \bibinfo {pages} {7750} (\bibinfo {year} {2019})}\BibitemShut {NoStop}%
\bibitem [{\citenamefont {Rau}\ \emph {et~al.}(2014)\citenamefont {Rau},
  \citenamefont {Baumann}, \citenamefont {Rusponi}, \citenamefont {Donati},
  \citenamefont {Stepanow}, \citenamefont {Gragnaniello}, \citenamefont
  {Dreiser}, \citenamefont {Piamonteze}, \citenamefont {Nolting}, \citenamefont
  {Gangopadhyay} \emph {et~al.}}]{rau2014reaching}%
  \BibitemOpen
  \bibfield  {author} {\bibinfo {author} {\bibfnamefont {I.~G.}\ \bibnamefont
  {Rau}}, \bibinfo {author} {\bibfnamefont {S.}~\bibnamefont {Baumann}},
  \bibinfo {author} {\bibfnamefont {S.}~\bibnamefont {Rusponi}}, \bibinfo
  {author} {\bibfnamefont {F.}~\bibnamefont {Donati}}, \bibinfo {author}
  {\bibfnamefont {S.}~\bibnamefont {Stepanow}}, \bibinfo {author}
  {\bibfnamefont {L.}~\bibnamefont {Gragnaniello}}, \bibinfo {author}
  {\bibfnamefont {J.}~\bibnamefont {Dreiser}}, \bibinfo {author} {\bibfnamefont
  {C.}~\bibnamefont {Piamonteze}}, \bibinfo {author} {\bibfnamefont
  {F.}~\bibnamefont {Nolting}}, \bibinfo {author} {\bibfnamefont
  {S.}~\bibnamefont {Gangopadhyay}}, \emph {et~al.},\ }\bibfield  {title}
  {\bibinfo {title} {Reaching the magnetic anisotropy limit of a 3$d$ metal
  atom},\ }\href@noop {} {\bibfield  {journal} {\bibinfo  {journal} {Science}\
  }\textbf {\bibinfo {volume} {344}},\ \bibinfo {pages} {988} (\bibinfo {year}
  {2014})}\BibitemShut {NoStop}%
\bibitem [{\citenamefont {Goodenough}(1963)}]{GKA}%
  \BibitemOpen
  \bibfield  {author} {\bibinfo {author} {\bibfnamefont {J.~B.}\ \bibnamefont
  {Goodenough}},\ }\href@noop {} {\emph {\bibinfo {title} {Magnetism and the
  Chemical Bond}}}\ (\bibinfo  {publisher} {Interscience-Wiley, New York},\
  \bibinfo {year} {1963})\BibitemShut {NoStop}%
\bibitem [{\citenamefont {Kim}\ \emph {et~al.}(2019{\natexlab{b}})\citenamefont
  {Kim}, \citenamefont {Kim}, \citenamefont {Kim}, \citenamefont {Kang},
  \citenamefont {Shin}, \citenamefont {Lee}, \citenamefont {Min},\ and\
  \citenamefont {Park}}]{kim2019exploitable}%
  \BibitemOpen
  \bibfield  {author} {\bibinfo {author} {\bibfnamefont {J.}~\bibnamefont
  {Kim}}, \bibinfo {author} {\bibfnamefont {K.-W.}\ \bibnamefont {Kim}},
  \bibinfo {author} {\bibfnamefont {B.}~\bibnamefont {Kim}}, \bibinfo {author}
  {\bibfnamefont {C.-J.}\ \bibnamefont {Kang}}, \bibinfo {author}
  {\bibfnamefont {D.}~\bibnamefont {Shin}}, \bibinfo {author} {\bibfnamefont
  {S.-H.}\ \bibnamefont {Lee}}, \bibinfo {author} {\bibfnamefont {B.-C.}\
  \bibnamefont {Min}},\ and\ \bibinfo {author} {\bibfnamefont {N.}~\bibnamefont
  {Park}},\ }\bibfield  {title} {\bibinfo {title} {Exploitable magnetic
  anisotropy of the two-dimensional magnet {CrI$_3$}},\ }\href@noop {}
  {\bibfield  {journal} {\bibinfo  {journal} {Nano Lett.}\ }\textbf {\bibinfo
  {volume} {20}},\ \bibinfo {pages} {929} (\bibinfo {year}
  {2019}{\natexlab{b}})}\BibitemShut {NoStop}%
\bibitem [{\citenamefont {Moriya}(1960)}]{pr1960}%
  \BibitemOpen
  \bibfield  {author} {\bibinfo {author} {\bibfnamefont {T.}~\bibnamefont
  {Moriya}},\ }\bibfield  {title} {\bibinfo {title} {Anisotropic superexchange
  interaction and weak ferromagnetism},\ }\href
  {https://doi.org/10.1103/PhysRev.120.91} {\bibfield  {journal} {\bibinfo
  {journal} {Phys. Rev.}\ }\textbf {\bibinfo {volume} {120}},\ \bibinfo {pages}
  {91} (\bibinfo {year} {1960})}\BibitemShut {NoStop}%
\bibitem [{\citenamefont {Liang}\ \emph {et~al.}(2020)\citenamefont {Liang},
  \citenamefont {Wang}, \citenamefont {Du}, \citenamefont {Hallal},
  \citenamefont {Garcia}, \citenamefont {Chshiev}, \citenamefont {Fert},\ and\
  \citenamefont {Yang}}]{MXY}%
  \BibitemOpen
  \bibfield  {author} {\bibinfo {author} {\bibfnamefont {J.}~\bibnamefont
  {Liang}}, \bibinfo {author} {\bibfnamefont {W.}~\bibnamefont {Wang}},
  \bibinfo {author} {\bibfnamefont {H.}~\bibnamefont {Du}}, \bibinfo {author}
  {\bibfnamefont {A.}~\bibnamefont {Hallal}}, \bibinfo {author} {\bibfnamefont
  {K.}~\bibnamefont {Garcia}}, \bibinfo {author} {\bibfnamefont
  {M.}~\bibnamefont {Chshiev}}, \bibinfo {author} {\bibfnamefont
  {A.}~\bibnamefont {Fert}},\ and\ \bibinfo {author} {\bibfnamefont
  {H.}~\bibnamefont {Yang}},\ }\bibfield  {title} {\bibinfo {title} {Very large
  {Dzyaloshinskii-Moriya} interaction in two-dimensional janus manganese
  dichalcogenides and its application to realize skyrmion states},\ }\href
  {https://doi.org/10.1103/PhysRevB.101.184401} {\bibfield  {journal} {\bibinfo
   {journal} {Phys. Rev. B}\ }\textbf {\bibinfo {volume} {101}},\ \bibinfo
  {pages} {184401} (\bibinfo {year} {2020})}\BibitemShut {NoStop}%
\bibitem [{\citenamefont {Ma}\ \emph {et~al.}(2018{\natexlab{a}})\citenamefont
  {Ma}, \citenamefont {Yu}, \citenamefont {Tang}, \citenamefont {Li},
  \citenamefont {He}, \citenamefont {Shi}, \citenamefont {Wang},\ and\
  \citenamefont {Li}}]{dmi_prl2018}%
  \BibitemOpen
  \bibfield  {author} {\bibinfo {author} {\bibfnamefont {X.}~\bibnamefont
  {Ma}}, \bibinfo {author} {\bibfnamefont {G.}~\bibnamefont {Yu}}, \bibinfo
  {author} {\bibfnamefont {C.}~\bibnamefont {Tang}}, \bibinfo {author}
  {\bibfnamefont {X.}~\bibnamefont {Li}}, \bibinfo {author} {\bibfnamefont
  {C.}~\bibnamefont {He}}, \bibinfo {author} {\bibfnamefont {J.}~\bibnamefont
  {Shi}}, \bibinfo {author} {\bibfnamefont {K.~L.}\ \bibnamefont {Wang}},\ and\
  \bibinfo {author} {\bibfnamefont {X.}~\bibnamefont {Li}},\ }\bibfield
  {title} {\bibinfo {title} {Interfacial dzyaloshinskii-moriya interaction:
  Effect of $5d$ band filling and correlation with spin mixing conductance},\
  }\href {https://doi.org/10.1103/PhysRevLett.120.157204} {\bibfield  {journal}
  {\bibinfo  {journal} {Phys. Rev. Lett.}\ }\textbf {\bibinfo {volume} {120}},\
  \bibinfo {pages} {157204} (\bibinfo {year} {2018}{\natexlab{a}})}\BibitemShut
  {NoStop}%
\bibitem [{\citenamefont {Ma}\ \emph {et~al.}(2018{\natexlab{b}})\citenamefont
  {Ma}, \citenamefont {Yu}, \citenamefont {Tang}, \citenamefont {Li},
  \citenamefont {He}, \citenamefont {Shi}, \citenamefont {Wang},\ and\
  \citenamefont {Li}}]{PRL_5d}%
  \BibitemOpen
  \bibfield  {author} {\bibinfo {author} {\bibfnamefont {X.}~\bibnamefont
  {Ma}}, \bibinfo {author} {\bibfnamefont {G.}~\bibnamefont {Yu}}, \bibinfo
  {author} {\bibfnamefont {C.}~\bibnamefont {Tang}}, \bibinfo {author}
  {\bibfnamefont {X.}~\bibnamefont {Li}}, \bibinfo {author} {\bibfnamefont
  {C.}~\bibnamefont {He}}, \bibinfo {author} {\bibfnamefont {J.}~\bibnamefont
  {Shi}}, \bibinfo {author} {\bibfnamefont {K.~L.}\ \bibnamefont {Wang}},\ and\
  \bibinfo {author} {\bibfnamefont {X.}~\bibnamefont {Li}},\ }\bibfield
  {title} {\bibinfo {title} {Interfacial {Dzyaloshinskii-Moriya} interaction:
  Effect of $5d$ band filling and correlation with spin mixing conductance},\
  }\href {https://doi.org/10.1103/PhysRevLett.120.157204} {\bibfield  {journal}
  {\bibinfo  {journal} {Phys. Rev. Lett.}\ }\textbf {\bibinfo {volume} {120}},\
  \bibinfo {pages} {157204} (\bibinfo {year} {2018}{\natexlab{b}})}\BibitemShut
  {NoStop}%
\bibitem [{\citenamefont {Yuan}\ \emph {et~al.}(2020)\citenamefont {Yuan},
  \citenamefont {Yang}, \citenamefont {Cai}, \citenamefont {Wu}, \citenamefont
  {Chen}, \citenamefont {Yan},\ and\ \citenamefont {Shen}}]{Janus_1}%
  \BibitemOpen
  \bibfield  {author} {\bibinfo {author} {\bibfnamefont {J.}~\bibnamefont
  {Yuan}}, \bibinfo {author} {\bibfnamefont {Y.}~\bibnamefont {Yang}}, \bibinfo
  {author} {\bibfnamefont {Y.}~\bibnamefont {Cai}}, \bibinfo {author}
  {\bibfnamefont {Y.}~\bibnamefont {Wu}}, \bibinfo {author} {\bibfnamefont
  {Y.}~\bibnamefont {Chen}}, \bibinfo {author} {\bibfnamefont {X.}~\bibnamefont
  {Yan}},\ and\ \bibinfo {author} {\bibfnamefont {L.}~\bibnamefont {Shen}},\
  }\bibfield  {title} {\bibinfo {title} {Intrinsic skyrmions in monolayer
  {Janus} magnets},\ }\href {https://doi.org/10.1103/PhysRevB.101.094420}
  {\bibfield  {journal} {\bibinfo  {journal} {Phys. Rev. B}\ }\textbf {\bibinfo
  {volume} {101}},\ \bibinfo {pages} {094420} (\bibinfo {year}
  {2020})}\BibitemShut {NoStop}%
\bibitem [{\citenamefont {Fert}\ and\ \citenamefont {Levy}(1980)}]{PRL1980}%
  \BibitemOpen
  \bibfield  {author} {\bibinfo {author} {\bibfnamefont {A.}~\bibnamefont
  {Fert}}\ and\ \bibinfo {author} {\bibfnamefont {P.~M.}\ \bibnamefont
  {Levy}},\ }\bibfield  {title} {\bibinfo {title} {Role of anisotropic exchange
  interactions in determining the properties of spin-glasses},\ }\href
  {https://doi.org/10.1103/PhysRevLett.44.1538} {\bibfield  {journal} {\bibinfo
   {journal} {Phys. Rev. Lett.}\ }\textbf {\bibinfo {volume} {44}},\ \bibinfo
  {pages} {1538} (\bibinfo {year} {1980})}\BibitemShut {NoStop}%
\bibitem [{\citenamefont {Sandratskii}(2017)}]{dmi2017PRB}%
  \BibitemOpen
  \bibfield  {author} {\bibinfo {author} {\bibfnamefont {L.~M.}\ \bibnamefont
  {Sandratskii}},\ }\bibfield  {title} {\bibinfo {title} {Insight into the
  {Dzyaloshinskii-Moriya} interaction through first-principles study of chiral
  magnetic structures},\ }\href {https://doi.org/10.1103/PhysRevB.96.024450}
  {\bibfield  {journal} {\bibinfo  {journal} {Phys. Rev. B}\ }\textbf {\bibinfo
  {volume} {96}},\ \bibinfo {pages} {024450} (\bibinfo {year}
  {2017})}\BibitemShut {NoStop}%
\bibitem [{\citenamefont {Zhang}\ \emph {et~al.}(2022)\citenamefont {Zhang},
  \citenamefont {Li}, \citenamefont {Wu}, \citenamefont {Wang}, \citenamefont
  {Zhao},\ and\ \citenamefont {Gao}}]{PRB2022}%
  \BibitemOpen
  \bibfield  {author} {\bibinfo {author} {\bibfnamefont {F.}~\bibnamefont
  {Zhang}}, \bibinfo {author} {\bibfnamefont {X.}~\bibnamefont {Li}}, \bibinfo
  {author} {\bibfnamefont {Y.}~\bibnamefont {Wu}}, \bibinfo {author}
  {\bibfnamefont {X.}~\bibnamefont {Wang}}, \bibinfo {author} {\bibfnamefont
  {J.}~\bibnamefont {Zhao}},\ and\ \bibinfo {author} {\bibfnamefont
  {W.}~\bibnamefont {Gao}},\ }\bibfield  {title} {\bibinfo {title} {Strong
  dzyaloshinskii-moriya interaction in monolayer ${\mathrm{cri}}_{3}$ on metal
  substrates},\ }\href {https://doi.org/10.1103/PhysRevB.106.L100407}
  {\bibfield  {journal} {\bibinfo  {journal} {Phys. Rev. B}\ }\textbf {\bibinfo
  {volume} {106}},\ \bibinfo {pages} {L100407} (\bibinfo {year}
  {2022})}\BibitemShut {NoStop}%
\end{thebibliography}%

\end{document}